\documentclass[aps,prl,showpacs,amsmath,amssymb,twocolumn]{revtex4-1}
\usepackage{graphicx}

\begin{document}

\title{Disordered bosons in one dimension: from weak to strong randomness criticality}

\author{Fawaz Hrahsheh and  Thomas Vojta}
\affiliation{Department of Physics, Missouri University of Science and Technology, Rolla, MO 65409}

\date{\today}

\begin{abstract}
We investigate the superfluid-insulator quantum phase transition of one-dimensional bosons with
off-diagonal disorder by means of large-scale Monte-Carlo simulations. For weak disorder, we find
the transition to be in the same universality class as the superfluid-Mott insulator transition
of the clean system. The nature of the transition changes for stronger disorder. Beyond a critical
disorder strength, we find nonuniversal, disorder-dependent critical behavior. We compare our results
to recent perturbative and strong-disorder renormalization group predictions. We also discuss
experimental implications as well as extensions of our results to other systems.
\end{abstract}


\maketitle


Bosonic many-particle systems can undergo quantum phase transitions between superfluid and
localized ground states due to interactions and lattice effects. These superfluid-insulator
transitions occur in a wide variety of experimental systems ranging from helium in porous media,
Josephson junction arrays, and granular superconductors to ultracold atomic gases
\cite{CHSTR83,CBMWR88,HavilandLiuGoldman89,HebardPaalanen90,BezryadinLauTinkham00,GMEHB02,DZSZL03,FLGFI07}.
In many of these applications, the bosons are subject to quenched disorder or randomness.
Understanding the effects of disorder on the superfluid-insulator transition and on the
resulting insulating phases is thus a prime question.

The case of one space dimension is especially interesting because the superfluid phase is rather subtle
and displays
quasi-long-range order instead of true long-range order. Moreover, the Anderson localization
scenario for non-interacting bosons suggests that disorder becomes more important
with decreasing dimensionality.

Giarmarchi and Schulz \cite{GiamarchiSchulz87,*GiamarchiSchulz88} studied the influence of
weak disorder on the interacting superfluid by means of a perturbative renormalization group
analysis. They found the superfluid-insulator transition to be of Kosterlitz-Thouless
(KT) type \cite{KosterlitzThouless73}, with universal critical exponents and a universal
value of the Luttinger parameter $g=\pi\sqrt{\rho_s \kappa}$ at criticality ($\rho_s$ denotes
the superfluid stiffness and $\kappa$ the compressibility). This analysis was recently
extended to second order in the disorder strength, with unchanged conclusion \cite{RPDG12}.

A different scenario emerges, however, from the real-space strong-disorder renormalization
group approach. In a series of papers \cite{AKPR04,*AKPR08,*AKPR10}, Altman et al.\
studied one-dimensional interacting lattice bosons in various types of disorder.
In all cases, they found that the superfluid-insulator transition is characterized by KT-like scaling of
lengths and times, but it occurs at a nonuniversal, disorder-dependent value of the Luttinger
parameter. The transition is thus in a different universality class than the weak-disorder
transition \cite{GiamarchiSchulz87,GiamarchiSchulz88}. However, Monte-Carlo
simulations \cite{BalabayanProkofevSvistunov05} did not find any evidence in favor of the strong-disorder critical point.

In view of these seemingly incompatible results, it is important to determine whether or not
both types of superfluid-insulator critical points indeed exist in systems of interacting
disordered bosons in one dimension. Moreover, it is important to study whether they can be reached for
realistic disorder strengths.

\begin{figure}
\includegraphics[width=8.cm]{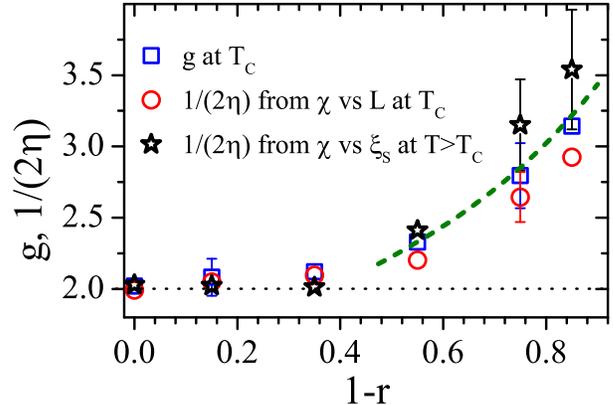}
\caption{(Color online) Critical Luttinger parameter $g$ and exponent $\eta$ [plotted as $1/(2\eta)$] of the superfluid-insulator transition
         as functions of the disorder strength $1-r$. The critical behavior appears universal for weak disorder
         while it becomes disorder-dependent for strong disorder. The lines are guides to the eye only.}
\label{fig:summary}
\end{figure}

In this Letter, we employ large-scale Monte-Carlo simulations to address these questions.
We focus on the case of off-diagonal
disorder at large commensurate filling; other types of disorder will be discussed
in the conclusions. Our results can be summarized as follows (see Fig.\ \ref{fig:summary}).
For weak disorder, we find a KT critical point in the universality class of the
clean (1+1) dimensional XY model, with universal exponents and a universal value of the Luttinger parameter
at the transition. This agrees with the predictions of the perturbative renormalization group.
If the disorder strength is increased beyond a threshold value, the nature of the transition changes.
While the scaling of length and time scales remains KT-like, the critical exponents and the
Luttinger parameter become non-universal, in agreement
with the strong-disorder scenario \cite{AKPR04,*AKPR08,*AKPR10}.
In the remainder of this Letter, we explain how these results were obtained, we discuss their generality
as well as implications for experiment.


The starting point is the disordered one-dimensional quantum rotor Hamiltonian

\begin{equation}
 H = -\sum_j J_j \cos(\hat \phi_{j+1}-\hat \phi_j) + \frac 1 2 \sum_j U_j (\hat n_j - \bar n_j)^2
\label{eq:Hamiltonian}
\end{equation}
which represents, e.g., a chain of superfluid grains with Josephson couplings $J_j$,
charging energies $U_j$ and offset charges $\bar n_j$.  $\hat n_j$ is the charge on grain $j$
and $\hat \phi_j$ is the phase of the superfluid order parameter. This model has a superfluid ground state
if the Josephson couplings dominate. With increasing charging energies it undergoes a quantum phase
transition to an insulating ground state.
In addition to Josephson junction arrays,
the Hamiltonian (\ref{eq:Hamiltonian}) describes a wide variety of other systems
that undergo superfluid-insulator transitions.

Within the strong-disorder approach \cite{AKPR04,*AKPR08,*AKPR10}, the type of insulator depends on the
symmetry properties of the offset charge distribution. In contrast, these details were found unimportant
at the critical point. In the following, we therefore focus on purely off-diagonal disorder,
$\bar n_j =0$. In this case, the Hamiltonian (\ref{eq:Hamiltonian}) can be mapped onto a classical
$(1+1)$-dimensional XY model \cite{WSGY94}
\begin{equation}
 H_{\rm cl} = -\sum_{j,\tau}\left [ J^s_j \cos(\phi_{j+1,\tau}-\phi_{j,\tau}) + J^t_j \cos(\phi_{j,\tau+1}-\phi_{j,\tau})
 \right ]
\label{eq:Hcl}
\end{equation}
where $j$ and $\tau$ index the lattice sites in the space and time-like directions, respectively.
The coupling constants $J^s_j/T$ and $J^t_j/T$ are determined by the parameters of the original
Hamiltonian (\ref{eq:Hamiltonian}) with $T$ being an effective
``classical'' temperature, not equal to the real physical temperature which is zero. In the following,
we fix $J^s_j$ and $J^t_j$ and drive the XY model (\ref{eq:Hcl}) through the transition
by tuning $T$. The interactions $J^s_j$ and/or $J^t_j$ are independent random variables
drawn from probability distributions $P^s(J^s)$ and $P^t(J^t)$. They depend on the space coordinate $j$ only;
this means the disorder
is columnar (perfectly correlated in time direction).

To determine the critical behavior of the classical XY model (\ref{eq:Hcl}), we performed large-scale Monte-Carlo
simulations using the efficient Wolff cluster algorithm \cite{Wolff89}. We studied square lattices
with linear sizes up to $L= 3200$ and averaged the results over large numbers (200 to 3000, depending on $L$)
of disorder realizations. Each sample was equilibrated using 200 to 400 Monte-Carlo sweeps, i.e., total spin flips per site.
(The actual equilibration times both above and at the critical temperature $T_c$ did not exceed about 20 sweeps.)
During the measurement period of 5000 to 30000 sweeps, we calculated observables such as
specific heat, magnetization, susceptibility, spin-wave stiffness as well as correlation functions.
In most simulations, we employed a uniform $J^s_j=1$ and drew the $J^t_j$ from a
binary probability distribution
\begin{equation}
P^t(J^t)= c \delta(J^t - r) + (1-c) \delta (J^t-1)~.
\label{eq:distribution}
\end{equation}
Here, $c$ is the concentration of weak bonds which we fixed at $c=0.8$. The disorder strength was tuned
by changing the value $r$ of the weak bonds. In addition to the clean case $r=1$ (which corresponds to
the pure superfluid-Mott insulator transition), we used $r=0.85, 0.65, 0.45, 0.25$, and
0.15. We also carried out test calculations with random $J^s$. All simulations were performed
on the Pegasus Cluster at Missouri S\&T, using about 400,000 CPU hours

\begin{figure}
\includegraphics[width=8.4cm]{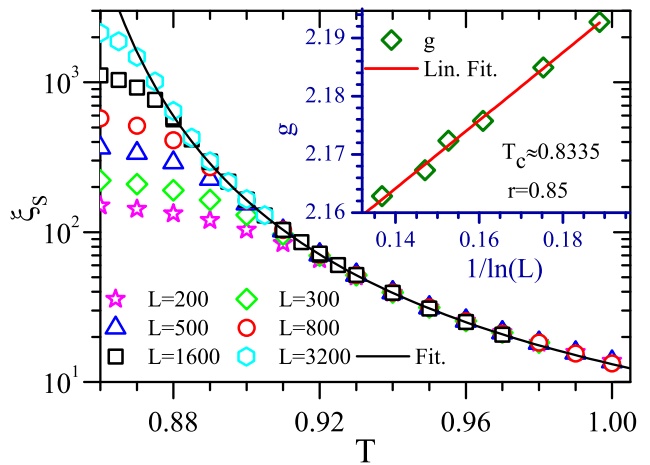}
\caption{(Color online) Spatial correlation length $\xi_s$ vs. temperature $T$
for disorder strength $r=0.85$ and system sizes $L=200$ to 3200. The solid line is
a fit to the KT form (\ref{eq:xi_KT}). Inset: Luttinger parameter $g$ at $T_c$ vs.
system size $L$.}
\label{fig:xi}
\end{figure}

We now turn to the results. To find $T_c$ for each disorder strength $r$,
we analyzed the behavior of the correlation length $\xi_s$ (in the space-like direction indexed
by $j$). It is calculated, as usual, from the second moment of the disorder-averaged correlation function.
In the high-temperature phase but close to the transition, $\xi_s$ is expected
to follow the form
\begin{equation}
\xi_s = A \exp\left[B(T-T_c)^{-1/2}\right]
\label{eq:xi_KT}
\end{equation}
both in the clean KT universality class \cite{KosterlitzThouless73} and
in the strong-disorder scenario \cite{AKPR04,*AKPR08,*AKPR10}.
$A$ and $B$ are non-universal constants. For all disorder strength, our data
follow this prediction with high accuracy, see Fig.\ \ref{fig:xi}
for an example.
We extract $T_c$ from fits of the data to (\ref{eq:xi_KT}) restricted to
$\xi_s>10$ to be in the critical region but $\xi_s<L/10$
to avoid finite-size effects. In the clean case ($r=1$), we obtain $T_c=0.8924$
in excellent agreement with high-precision values in the literature \cite{Hasenbusch05}
\footnote{The remaining small difference can be attributed to logarithmic corrections
to (\ref{eq:xi_KT}) which we did not account for.}.

In addition to the correlation length $\xi_s$ in the space-like direction, we also
studied the correlation length $\xi_t$ in the time-like direction. We found
$\xi_t \propto  \xi_s$ for all disorder strengths which implies a dynamical exponent
of $z=1$.

\begin{figure}
\includegraphics[width=8.cm]{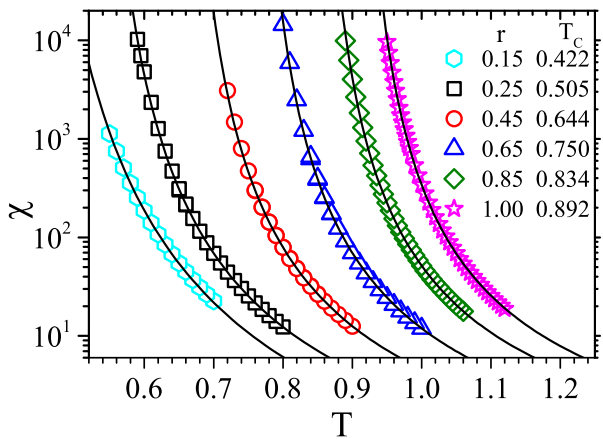}
\caption{(Color online) Susceptibility $\chi$ vs. temperature $T$
for several disorder strengths. The maximum system sizes are at least $L=1500$.
The solid lines are fits to the KT form (\ref{eq:chi_KT}). The
resulting estimates of $T_c$ are listed in the legend.}
\label{fig:chi}
\end{figure}

The order parameter susceptibility $\chi$ can be analyzed analogously. In the high-temperature phase
close to the transition, it is predicted to behave as
\begin{equation}
\chi \propto \xi_s^{2-\eta} \propto  \exp\left[D(T-T_c)^{-1/2}\right]~.
\label{eq:chi_KT}
\end{equation}
Here, $\eta$ is the correlation function critical exponent and
$D=(2-\eta) B$. Figure \ref{fig:chi} shows that the data for all disorder
strengths $r$ follow this prediction with high accuracy.
The critical temperatures extracted from the corresponding fits are listed in the
legend of the figure. Their values have small statistical errors ranging from about
$3\times 10^{-4}$ for the weak disorder cases to $2\times 10^{-3}$ for strong
disorder. The systematic errors due to corrections to the leading scaling form (\ref{eq:chi_KT})
are somewhat larger. We estimate them from the robustness of the fit against changing
the fit interval. This yields systematic errors ranging from about $5 \times 10^{-3}$
for weak disorder to $2\times 10^{-2}$ for strong disorder. Within these errors
the critical temperatures extracted from $\chi$ agree well with
those from the correlation lengths.

\begin{figure}
\includegraphics[width=8.2cm]{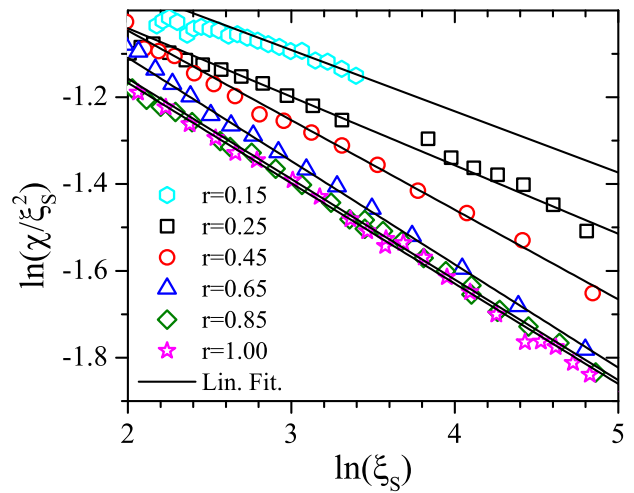}
\caption{(Color online) $\ln(\chi/\xi^2_s)$ vs. $\ln(\xi_s)$
for several disorder strengths and maximum system size $L\ge 1500$
($L=500$ for $r=0.15$). The solid lines are
linear fits; their slopes give $-\eta$.}
\label{fig:chi_xi}
\end{figure}

Equation (\ref{eq:chi_KT}) suggests a direct way to measure the exponent
$\eta$: if one plots $\ln(\chi/\xi_s^2)$ vs.\ $\ln(\xi_s)$, the data should be
on a straight line with slope $-\eta$.  Figure \ref{fig:chi_xi} presents this
analysis for different disorder strengths.
In the clean case, $r=1$, we find $\eta=0.243$ in good agreement with the exact
value 1/4 \cite{KosterlitzThouless73}.
The weak-disorder curves ($r=0.85$ and 0.65) are parallel to the clean one
within their statistical errors. Fits in the range $20<\xi_s<L/10$ give
exponents $\eta$ close to $1/4$. In contrast, the strong-disorder curves ($r=0.45$, 0.25, 0.15)
are less steep, resulting in smaller $\eta$. They
are also noisier which leads to larger error bars. All $\eta$ values
are shown in Fig.\ \ref{fig:summary}.
They provide evidence for universal critical behavior (in the clean 2D XY
universality class) for weak disorder but nonuniversal behavior for strong disorder.

In addition to simulations in the high-temperature phase, we also studied the finite-size
scaling properties of observables right at the critical temperature $T_c$.
Let us first consider the  Luttinger parameter $g=\pi\sqrt{\rho_s \kappa}$. Under the
quantum-to-classical mapping \cite{WSGY94}, the compressibility $\kappa$ of the
quantum rotor Hamiltonian (\ref{eq:Hamiltonian}) maps onto the spin-wave stiffness
$\rho_t$ in the time-like direction of the classical XY model (\ref{eq:Hcl}). In
our simulations, the Luttinger parameter is thus given by
\begin{equation}
g= (\pi/T)\sqrt{\rho_s \rho_t}~.
\label{eq:g}
\end{equation}
The stiffnesses $\rho_s$ and $\rho_t$ are not calculated by actually applying twisted boundary conditions
during the simulation but by using the relation given by Teitel and Jayaprakash \cite{TeitelJayaprakash83}
(for a derivation see, e.g., Ref.\ \cite{HrahshehBarghathiVojta11}).

Within KT theory, the Luttinger parameter
close to the transition behaves as $g(T) = g(T_c) + a (T_c-T)^{1/2}$
where $a$ is a constant and $T\le T_c$. Together with (\ref{eq:xi_KT}), this suggests the
leading finite-size corrections to $g$ at $T_c$ to take the form
\begin{equation}
g(T_c,L) = g(T_c,\infty) + b/\ln(L)
\label{eq:g_L}
\end{equation}
where $b$ is another constant. Calculating the Luttinger parameter at $T_c$
for different system sizes and extrapolating using
(\ref{eq:g_L}) yields the infinite-system value  $g(T_c,\infty)$
\footnote{The extrapolation of $g$ to $L=\infty$ is nontrivial
as $g$ shows a singular temperature dependence and a jump to $0$
for $T>T_c$. The data must be in the critical region,
$|T-T_c| \lesssim [\ln (L/A)]^{-2}$, which appears to be fulfilled in our
case.}.
We performed this analysis for all disorder strengths $r$ and found that the $g$ vs.\ $1/\ln(L)$ data
indeed fall onto straight lines (the inset of Fig.\ \ref{fig:xi} shows an example). The resulting extrapolated values are
displayed in Fig.\ \ref{fig:summary}. For weak disorder ($r=0.85$ and 0.65),
the Luttinger parameters at $T_c$ agree with the clean value, $g=2$, within their
error bars (which are combinations of the statistical Monte-Carlo error and the uncertainty in $T_c$).
For stronger disorder ($r=0.45$, 0.25, 0.15), $g(T_c,\infty)$ takes larger,
disorder-dependent values.

\begin{figure}
\includegraphics[width=8.2cm]{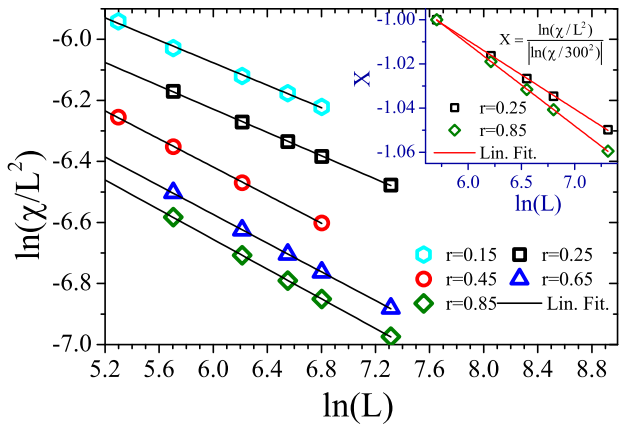}
\caption{(Color online) Susceptibility at $T_c$ plotted as $\ln(\chi/L^2_s)$ vs. $\ln(L)$
for several disorder strengths. The solid lines are
linear fits; their slopes give $-\eta$ (values are shown in Fig.\ \ref{fig:summary}) . The inset demonstrates the change of slope
with increasing $r$.}
\label{fig:chi_L}
\end{figure}

Finally, we turn to the finite-size behavior of the susceptibility at $T_c$.
According to finite-size scaling, the leading size-dependence should be of the
form
\begin{equation}
\chi(T_c,L) \sim L^{2-\eta}
\label{eq:chi_L}
\end{equation}
which provides another way to measure $\eta$. Figure \ref{fig:chi_L} shows
plots of $\ln(\chi/L^2)$ vs.\ $\ln(L)$ for all disorder strengths $r$.
For weak disorder ($r=0.85$ and 0.65), the resulting values of the exponent $\eta$
are again close to the clean value $1/4$. For larger disorder ($r=0.45$, 0.25, and 0.15),  we find
disorder-dependent values that roughly agree with those extracted in the high-temperature phase
(Fig.\ \ref{fig:chi_xi}).


In summary, we used large-scale Monte-Carlo simulations to investigate
the superfluid-insulator quantum phase transition of one-dimensional bosons
with off-diagonal disorder.
For weak disorder, our data provide evidence for a KT critical point in the universality class
of the clean (1+1) dimensional classical XY model, with universal critical
exponents $\eta=1/4$ and $z=1$ as well as a universal value $g=2$
of the critical Luttinger parameter. These results agree with the Harris
criterion \cite{Harris74} which predicts weak disorder to be an
irrelevant perturbation at the clean KT transition.
For stronger disorder, the universality class of the transition changes. It is
still of KT-type [$\xi_s$ and $\chi$ follow (\ref{eq:xi_KT}) and (\ref{eq:chi_KT})]
but the critical exponent $\eta$ and the critical
Luttinger parameter become disorder-dependent (non-universal)
\footnote{Fig.\ \ref{fig:summary} suggests that $g=1/(2\eta)$ not just at the clean
KT critical point but also at the strong disorder critical point. To the best of
our knowledge, the latter has not yet been established theoretically.}.
This agrees with the strong-disorder scenario \cite{AKPR04,*AKPR08,*AKPR10}.

The important question of whether the boundary between the weak and strong disorder
regimes is sharp or just a crossover cannot be finally decided
by means of our current numerical capabilities. The data in Fig.\ \ref{fig:summary}
would be compatible with both scenarios within their error bars.

Earlier Monte-Carlo simulations \cite{BalabayanProkofevSvistunov05} did not observe the
strong-disorder regime. We believe that the binary disorder used in \cite{BalabayanProkofevSvistunov05}
(equivalent to disorder in $J^s$ with $c=0.5$ and $r=0.33$ in our model) may not have been
sufficiently strong. In particular, $c=0.5$
is much less favorable for the formation of rare regions
than our $c=0.8$. To test this hypothesis, we
performed a few simulation using $c=0.5$ and $r=0.33$. They resulted in
critical behavior compatible with the clean 2D XY universality class, in agreement with
Ref.\ \cite{BalabayanProkofevSvistunov05} \footnote{Ref.\ \cite{BalabayanProkofevSvistunov05}
also studied power-law distributed interactions, but the results
showed significant finite-size effects.}.

It is interesting to ask whether the different critical behaviors in the weak and strong-disorder
regimes are accompanied by qualitative differences in the bulk phases. In particular,
are there two different insulating phases or are the weak and strong-disorder regimes
continuously connected? A detailed analysis of the insulating phase(s) will also shed
light on the mechanism that destroys the superfluid stiffness above $T_c$. Is it due
to the proliferation of single quantum phase slips as at a clean KT transition or due
to the formation of phase slip ``dipoles'' as suggested in Ref.\ \cite{AKPR04,*AKPR08,*AKPR10}?
Simulations to address these questions are under way.

All our explicit results are for off-diagonal disorder and large commensurate filling.
They do not directly apply to the generic dirty-boson problem with diagonal disorder
considered in \cite{GiamarchiSchulz87,*GiamarchiSchulz88}
\footnote{The critical value of $g$ in the perturbative theory
\cite{GiamarchiSchulz87} with diagonal disorder is 3/2 rather than 2.}.
Note, however, that the critical behavior does not depend on the disorder type
within the strong-disorder scenario \cite{AKPR04,*AKPR08,*AKPR10}.
 Simulating the generic case would require a different approach
(such as the link-current formulation \cite{WSGY94}) because the mapping onto a classical XY model is not
valid for diagonal disorder.

Finally, we turn to the experimental accessibility of the weak and strong-disorder regimes.
Our results show that the transition between them occurs at a moderate disorder strengths.
We therefore expect both regimes to be accessible in principle in experiments on systems such as ultracold atoms
or Josephson junction chains (see also Ref.\ \cite{VoskAltman12}).


We acknowledge discussions with Ehud Altman, David Pekker, Nikolay Prokof'ev, Gil Refael, and
Zoran Ristivojevic.
This work has been supported by the NSF under Grant Nos. DMR-0906566 and DMR-1205803.

\bibliographystyle{apsrev4-1}
\bibliography{../00Bibtex/rareregions}

\end{document}